\documentclass[%
 reprint,
superscriptaddress,
 amsmath,amssymb,
 aps,
showkeys,
]{revtex4-2}

\usepackage{graphicx,xcolor,soul,siunitx}
\graphicspath{{Figures/}}
\usepackage[version=3]{mhchem}
\usepackage{dcolumn}
\usepackage{bm}
\usepackage{cleveref}
\usepackage{multirow}

\newcommand\hil[1]{%
  \bgroup
  \hskip0pt\color{red!80!black}%
  #1%
  \egroup
}

\newif\ifhighlight
\highlightfalse
\ifhighlight
    
\else
    \renewcommand{\hil}[1]{#1}
\fi

\begin{document}

\preprint{APS/123-QED}

\title{Unsupervised deep learning framework for temperature-compensated damage assessment using ultrasonic guided waves on edge device}

\author{Pankhi Kashyap}
\affiliation{Department of Electrical Engineering, Indian Institute of Technology Bombay, Mumbai 400076, MH India}

\author{Kajal Shivgan}
\affiliation{Department of Electrical Engineering, Indian Institute of Technology Bombay, Mumbai 400076, MH India}

\author{Sheetal Patil}
\affiliation{Department of Electrical Engineering, Indian Institute of Technology Bombay, Mumbai 400076, MH India}

\author{Ramana Raja B.}
\affiliation{Department of Civil Engineering, Indian Institute of Technology Bombay, Mumbai 400076, MH India}

\author{Sagar Mahajan}
\affiliation{Department of Electrical Engineering, Indian Institute of Technology Bombay, Mumbai 400076, MH India}

\author{Sauvik Banerjee}
\affiliation{Department of Civil Engineering, Indian Institute of Technology Bombay, Mumbai 400076, MH India}

\author{Siddharth Tallur}
\email{stallur@ee.iitb.ac.in}
\affiliation{Department of Electrical Engineering, Indian Institute of Technology Bombay, Mumbai 400076, MH India}


\date{\today}

\begin{abstract}
Fueled by the rapid development of machine learning (ML) and greater access to cloud computing and graphics processing units (GPUs), various deep learning based models have been proposed for improving performance of ultrasonic guided wave structural health monitoring (GW-SHM) systems, especially to counter complexity and heterogeneity in data due to varying environmental factors (e.g., temperature) and types of damages. Such models typically comprise of millions of trainable parameters, and therefore add to cost of deployment due to requirements of cloud connectivity and processing, thus limiting the scale of deployment of GW-SHM. In this work,  we propose an alternative solution that leverages TinyML framework for development of light-weight ML models that could be directly deployed on embedded edge devices. The utility of our solution is illustrated by presenting an unsupervised learning framework for damage detection in honeycomb composite sandwich structure (HCSS) with disbond and delamination type of damages, validated using data generated by finite element (FE) simulations and experiments performed at various temperatures in the range \qtyrange{0}{90}{\degreeCelsius}. We demonstrate a fully-integrated solution using a Xilinx Artix\textregistered-7 FPGA for data acquisition and control, and edge-inference of damage.
\end{abstract}

                              
\maketitle

\section{Introduction}
Ultrasonic guided waves (GW) have the ability to propagate long distances without significant attenuation. GW based structural health monitoring (SHM) systems are highly sensitive to damages, and require minimal instrumentation of sensors on the structure to be monitored. Over the years, many studies has been reported on such systems, and a vast body of literature is available on GW SHM \cite{mitra2016guided}. However, unlike vibration condition based monitoring of rotating machinery \cite{randall2021vibration}, the industrial acceptance towards replacement of periodic nondestructive testing (NDT) with such permanently deployed SHM systems has been quite limited \cite{cawley2021development}. 
The primary hurdle in this regard is the impact of variations in environmental and operating conditions (EOCs) on GW propagation. The amplitude and group velocity of different GW modes have distinct temperature-dependent coefficients, and therefore assessing damages in presence of time-varying EOCs imposes a significant challenge \cite{mariani2020compensation}. 

Recently, data-driven damage assessment methods using machine learning (ML) techniques have been proposed to overcome this limitation, by identifying patterns hidden in the features computed from GW recordings \cite{bao2021machine}. Such reports include methods based on artificial neural networks (ANNs) and support vector machines (SVMs) \cite{agarwal2014lamb}, principal component analysis (PCA) \cite{liu2015robust}, singular value decomposition (SVD) \cite{clarke2010guided,figueiredo2011machine}, Gaussian mixture models (GMMs) \cite{ren2019gaussian,ren2019multi} etc.
On the other hand, deep learning techniques for GW based SHM eliminate the dependencies on domain knowledge for computation of features, by utilizing the deep learning models themselves for feature computation and classification \cite{RAUTELA2021114189, KHAN2019586, PANDEY2022108220}. Recently, \cite{sawant2022temperature} demonstrated a method which utilizes supervised convolutional neural networks (CNNs) to learn the features of GW signals acquired from healthy and damage structures for varying temperature conditions, combined with GMMs for damage index computation and localization.
Unlike such supervised methods that require labeled data collected for healthy as well as damaged structures, unsupervised learning approaches offer greater utility, as most practical applications of SHM systems will result in generation of significantly higher amount of data for healthy operating conditions as compared to damaged structure. Recently reported unsupervised learning demonstrations for GW-SHM include convolutional denoising autoencoder (AE) for temperature compensation \cite{rautela2021temperature} (albeit, without damage assessment) and deep AEs operating on time-domain GW data obtained on composite panels \cite{lee2022automated,sawant2023unsupervised}. Typical deep learning architectures containing convolutional layers result in millions of trainable parameters and require large amount of computational resources and take long time for training. Cloud based data storage and inference may be the most suitable option for meeting these requirements; however, the added cost of data transmission and storage, latency, and security and privacy aspects potentially limit their applications. Hence, there is a need for development of data driven unsupervised algorithms for damage assessment, that could be implemented in edge devices such as microcontrollers and field programmable gate arrays (FPGAs) used for data acquisition.

TinyML has emerged as a promising solution for seamless development of light-weight ML models on a variety of microcontrollers, resulting in a vast body of work on memory-efficient and low-power edge-implementation of ML applications \cite{sanchez2020tinyml,shi2016edge}. The ML models are typically trained on a workstation, and deployed on edge devices using device-specific libraries and open-source frameworks such as TensorFlow Lite Micro (TFLM) \cite{nguyen2019machine,david2021tensorflow}. While this enables deployment of models on low-cost microcontrollers with limited resources (e.g. single-board platforms such as Arduino), ML models typically require to be retrained in the field to accommodate drift in the sensor recordings, and therefore more powerful edge devices such as Raspberry Pi single board computer or FPGA platforms may be more suitable for practical applications \cite{kolcun2020case,malviya2022edge}. FPGAs offer design flexibility in GW-SHM applications, by allowing seamless reconfiguration of the ML model architecture and hardware implementation and acceleration of the multiply-accumulate (MAC) operations involved in data processing, without hardware redesign. The high speed clock offered (tens of MHz and higher) by most FPGAs also makes them suitable for interfacing with high speed digital to analog converters (DACs) and analog to digital converters (ADCs) for the GW transmitters and receivers,respectively. Aranguren et al. recently demonstrated an FPGA-based GW-SHM system for detection of progressive damage in aircraft structures \cite{aranguren2022methodology}. Malviya et al. demonstrated an FPGA-based implementation of light-weight convolutional AE model for anomaly detection in vibration-based monitoring, validated using the Airbus SAS helicopter accelerometer dataset \cite{malviya2022edge}. 
Thus, while embedded systems with edge-learning have been developed for a variety of applications, there have been no such reports for GW SHM to counter EOC variations. Moreover, models that leverage convolutional layers to learn features from vast amounts of training data are impractical due to low memory availability on such embedded systems.

\begin{figure}[!b]
    \centering{\includegraphics[width = \linewidth]{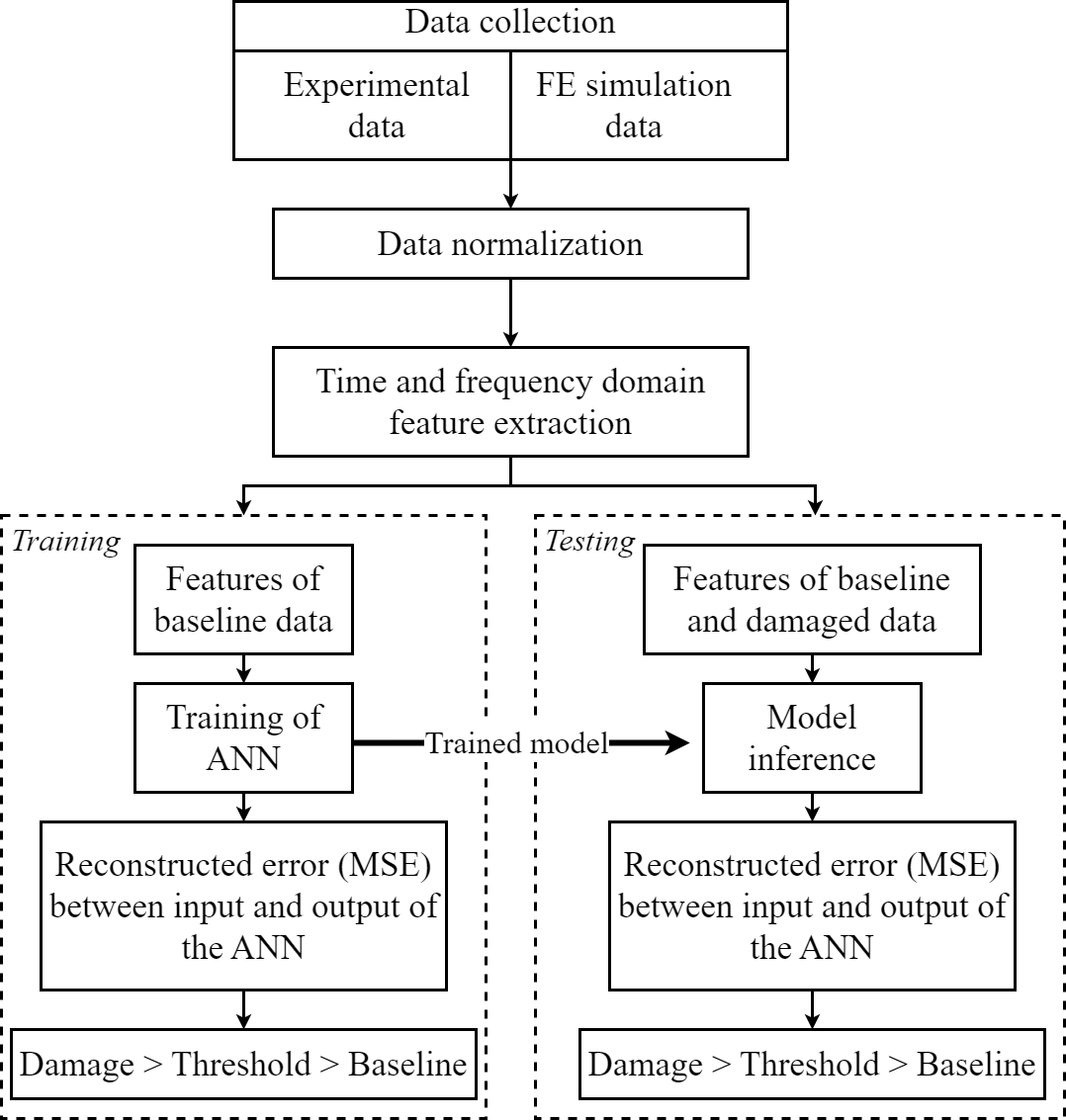}}
    \caption{Overview of the proposed methodology, described in section \ref{sec:overview}.
    }
    \label{fig:main_flow_chart}
\end{figure}

In this paper, we demonstrate a TinyML-enabled unsupervised learning framework for GW-SHM, implemented on the Xilinx Artix®-7 FPGA, to realize a fully-integrated SHM system for data acquisition and damage assessment (identification and localization). 
From a network of 8 PZT sensors covering an undamaged and two damaged portions of a panel made of honeycomb composite sandwich structure (HCSS), GW data are recorded at different temperatures from \qtyrange{0}{90}{\degreeCelsius}. A light-weight neural network is trained on features obtained from data collected on an undamaged panel, and is then tested on independently collected data from undamaged as well as damaged portions of the panel. The features comprise of various parameters obtained from time-domain GW signal recordings. The light-weight model is deployed using TensorFlow Lite framework on the MicroBlaze® reduced instruction set computer (RISC) processor core in the Xilinx Artix®-7 FPGA for damage assessment at the edge. The efficacy of this method is demonstrated using GW data from a network of PZT sensors on an HCSS panel with teflon release film (TRF) and lack of film adhesive (LFA) damages, data collected by experimental measures at various temperatures (from \qtyrange{0}{90}{\degreeCelsius}). A parametric study is also done using data generated by finite element (FE) simulations to see the model performance for detection of damages with different sizes. The mean square error (MSE) of the reconstructed signal is used as signal difference coefficient (SDC) for damage identification. The system presented in this work is a promising development towards fully-integrated data-driven GW-SHM solution.

\section{Methodology}
\subsection{Overview of proposed method}
\label{sec:overview}
The proposed damage assessment method broadly comprises of three steps:(i) feature extraction, (ii) training of unsupervised learning model, and (iii) edge implementation of trained model. 
Figure \ref{fig:main_flow_chart} shows the overview of our proposed methodology. First, time-series data were collected from a structure instrumented with PZT sensors, for healthy (undamaged) and damaged conditions. Data were generated using FE simulations and experimental setup for temperatures varying from \qtyrange{0}{90}{\degreeCelsius}. Next, features are computed from the signals,
based on the sensitivity to damage, computational cost and memory requirement associated with computing the features for edge implementation. Note that the features can be computed from time and frequency domain representations of the signals. In this work, we have only focused on the use of time-domain features, as they are simpler to compute and do not introduce any significant computation overhead on the embedded system. 
To study the performance of the model in presence of large errors, white noise and pink noise were added to the data to obtain signal to noise ratio (SNR) of \SI{20}{dB}. 
Next, an artificial neural network (ANN) was trained on the manually computed set of features, obtained using data collected for undamaged (baseline) structure under various temperature conditions. 
When the trained model is tested on data obtained from damaged panels, the paths containing damage located between the transmitter and receiver positions produce higher reconstruction error (mean squared error, MSE) for the model prediction, which in turn is used to identify damage. The trained model is deployed on a Xilinx Artix\textregistered-7 FPGA for damage inference using TensorFlow Lite framework operating in the MicroBlaze\textregistered~RISC processor in the FPGA. The FPGA is also used for data acquisition and feature computation, and thus serves as an end-to-end smart GW-SHM solution.

\subsection{Experimental data acquisition}
In this work, we have evaluated the proposed unsupervised learning method for damage assessment of a HCSS panel, which is an advanced composite structure widely used in aerospace, automotive and marine industries. 
The panel of lateral dimensions \qtyproduct{1x1.2}{m} was manufactured by sandwiching two composite face sheets with a \SI{12.7}{mm} thick aluminum honeycomb core between the face sheets. The face sheets comprised of six \SI{0.125}{mm} thick unidirectional layers with a layer-wise orientation of $[0, +60, -60, -60, +60, 0]$ and were separately cured and bonded to the core. Disbond was artificially created between the core and face sheet by removing the film adhesive in a region of dimensions \qtyproduct{30x30}{mm}, and delamination was created between the second and third layer of the face sheet adjacent to the core by inserting a thin layer of teflon release film of dimensions \qtyproduct{30x30}{mm}. The effect of temperature variation on the characteristics of GW signals for disbond (LFA) and delamination (TRF) damages in this structure was extensively characterized and reported in our previous work \cite{raja2023effect}. The amplitude and group velocity of the fundamental anti-symmetric (A$0$) mode were found to increase in presence of disbond and decrease in presence of face sheet delamination. However, the amplitude of A$0$ mode reduced linearly with increasing temperature for both healthy and damaged cases. The variation in amplitude of the A$0$ mode due to temperature makes it difficult to use this property to evaluate the presence of damage, especially in presence of large noise.

\begin{figure*}[!t]
    \centering{\includegraphics[width = \linewidth]{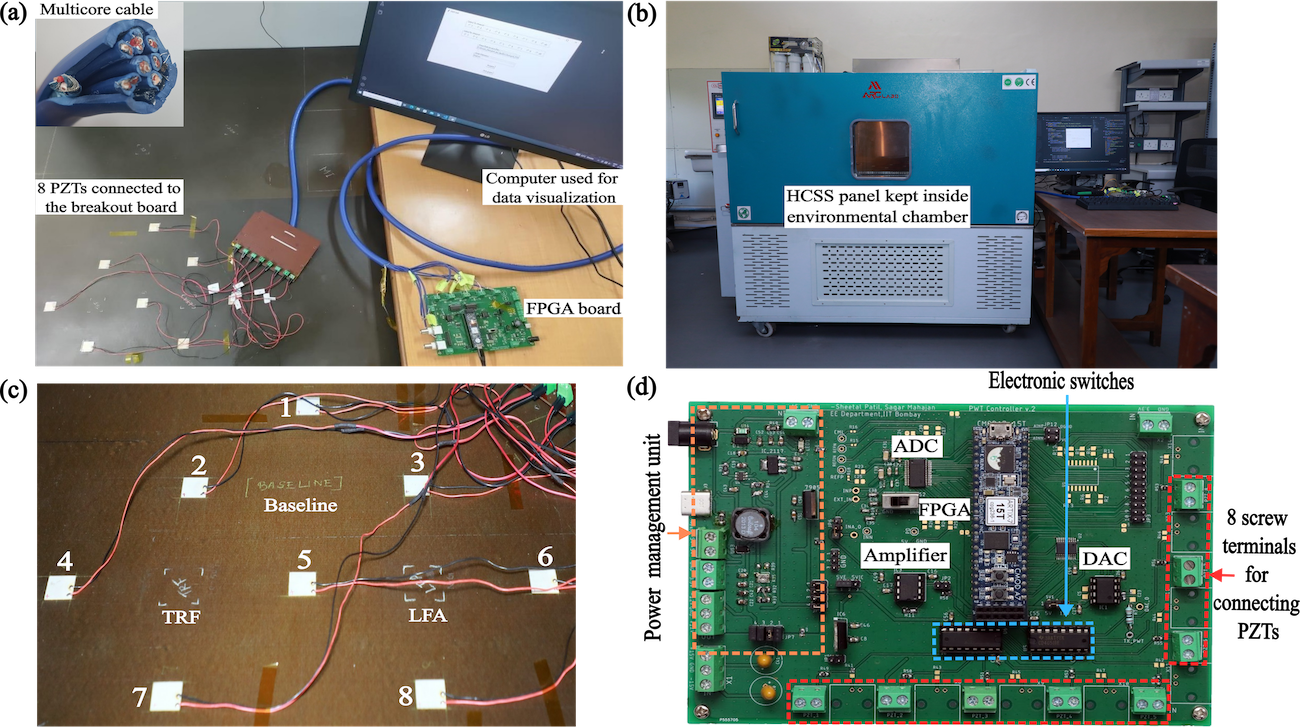}}
    \caption{Overview of the experimental setup. a) PZT network on HCSS structure connected to FPGA board using a shielded multicore cable. Inset: $8$ pairs of shielded cables in the multicore cable. (b) Environmental chamber used for temperature control. (c) HCSS panel showing positions of PZT transducers and damages (TRF and LFA) and undamaged (baseline) locations. (d) FPGA enabled embedded system for data acquisition and damage assessment.}
    \label{fig:setup}
\end{figure*}

Ultrasonic GW data from the structure were recorded by attaching a network of eight PZT-5H transducers on one of the face sheets of the structure (Figure \ref{fig:setup}(a)). Each transducer is of dimension \qtyproduct{30x30x0.4}{\milli\meter}. The transducers were placed in such a manner so that the distance between transmitter and receiver remains \SI{180}{mm} for any horizontal or vertical path in a unit cell.  
A multicore cable of length \SI{3}{m} with electromagnetic shielding was used to connect the PZTs with the FPGA board. Electromagnetic shielding prevents, or at the very least, reduces the coupling of undesired radiated electromagnetic energy in systems and cables. This is necessary in our experiment because the FPGA board is kept outside an environmental chamber (Arcade Scientific Instruments Pvt. Ltd., Figure \ref{fig:setup}(b)) used to record ultrasonic GW signals on the HCSS panel, and thus requires analog voltage signals from the PZTs to be connected with a long cable, that is susceptible to electromagnetic interference.
A closer view of the region of interest on the HCSS panel (Figure \ref{fig:setup}(c)) shows that the TRF damaged portion lies at the center of unit cell with PZTs numbered $2$, $7$ and $4$, $5$, and LFA damaged portion lies at the center of unit cell with PZTs numbered $3$, $8$ and $5$, $6$. Similarly, the region at the intersection of horizontal and vertical paths in unit cell with PZTs numbered $1$, $5$ and $2$, $3$ does not have any damage and is considered as baseline region. We collected time-series data from the above mentioned vertical and horizontal paths. Ultrasonic GW data were recorded for these paths by placing the HCSS panel in the environmental chamber and varying temperature from \qtyrange{0}{90}{\degreeCelsius}, in intervals of \SI{5}{\degreeCelsius}. 

Figure \ref{fig:setup}(d) shows the portable embedded system built for GW ultrasonic signal transduction, data acquisition, and processing. 
The system is based on a low-cost FPGA (Xilinx Artix\textregistered-7 XC7A15T-1CPG236C, Digilent Cmod A7-15 T module). The FPGA is interfaced with a $12$ bit, parallel input, multiplying digital to analog converter (DAC) — Texas Instruments DAC7821, with a transimpedance amplifier (Texas Instruments TL072) to generate the $5-$cycle Hanning pulse actuation voltage centered at \SI{75}{kHz} with peak-to-peak amplitude of \SI{10}{V}. The receiver PZT output is amplified using Texas Instruments INA128 instrumentation amplifier and digitized using Maxim Integrated MAX-1426 analog to digital converter (ADC) with $10$ bit resolution and \SI{10}{Msps} sampling rate. The printed circuit board (PCB) also consists of an on-board power supply module with voltage regulators for generating necessary supply rails for the amplifier and data converter ICs, so that the entire board can be powered with a USB type-C charger. Voltage controlled switches implemented with Texas Instruments CD4051 allow digital selection of the PZTs to be connected as transmitter and receiver to the signal chain. The system is capable of storing $4096$ samples obtained at \SI{10}{Msps} i.e., \SI{409.6}{\micro\second} for each channel in the memory (Block RAM) in the FPGA. A graphical user interface (GUI) created using Python is used for system control and configuration, and operating the experiment (selection of PZT channels, visualizing GW data and saving data to file). To incorporate uncertainties caused by variation in other environmental and operating conditions that one may expect in the field, we also added white noise and pink noise to the data to obtain SNR of \SI{20}{dB} \cite{sawant2022temperature} (note that the SNR of the data recorded with the system described above is $>$\SI{50}{dB}). Fifty noise-augmented copies were generated from each data recording.

\begin{figure}[!t]
    \centering
    \includegraphics[width=\linewidth]{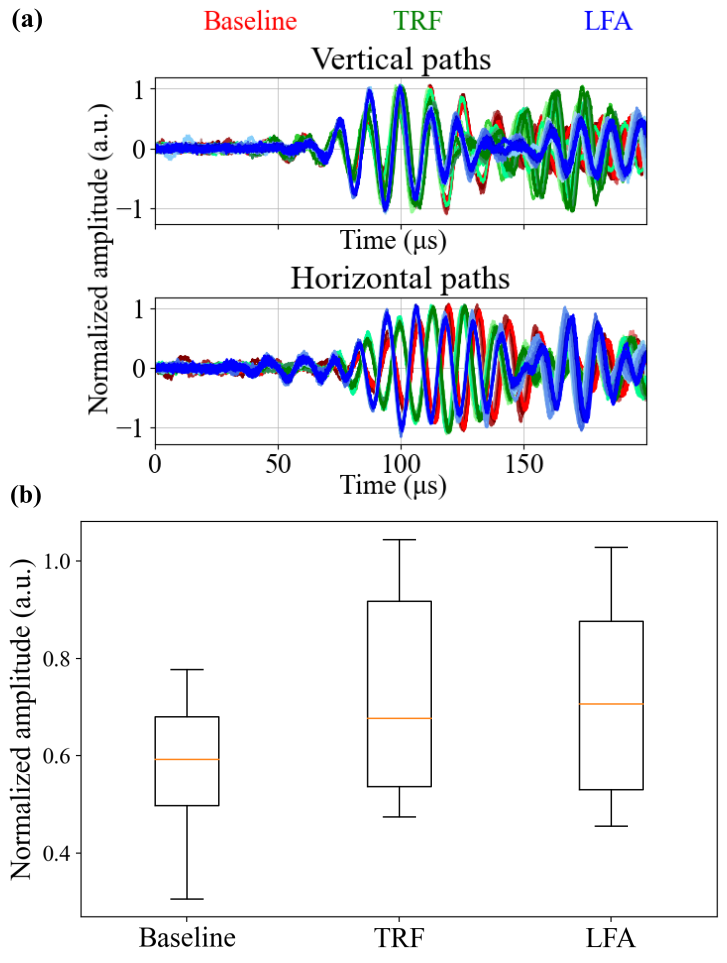}
    \caption{(a) Noise-augmented data series recorded from \qtyrange{0}{90}{\degreeCelsius}, for all horizontally and vertically oriented paths passing through baseline, TRF and LFA regions. Lighter shades of color represents lower temperature and darker shades represent progressively higher temperatures. (b) Variation in amplitude of A$0$ mode due to temperature, shown as distribution for baseline, TRF and LFA damages.}
    
    \label{fig:rgb_plot}
\end{figure}

Figure \ref{fig:rgb_plot}(a) shows a plot of representative noise-augmented data gathered at all temperatures for vertical and horizontal paths in all unit cells up to \SI{200}{\micro\second}.
Data acquired at higher temperatures are shown using deeper color shades, whereas data obtained at lower temperatures are indicated by lighter color shades. The variation in amplitude of the A$0$ mode due to temperature for vertical and horizontal paths in these noise-augmented data are more clearly seen in Figure \ref{fig:rgb_plot}(b). In addition to change in amplitude, the group velocity of the A$0$ mode also changes due to temperature, making it difficult to rely solely on analytical methods to assess damage. 

\subsection{Data generation through finite element method (FEM) simulations}
The experimental sample used in this study contained only a few damages, one each of TRF and LFA damage. Therefore, to study the sensitivity of the method proposed in this work for damage assessment for various sizes of defects, additional data were generated through FEM simulations.
Building a 3D model of the entire panel (lateral dimensions \qtyproduct{1x1.2}{m}) is computationally expensive and therefore impractical. Since the GW propagation paths used in our study are straight lines, we employed a 2D model to simplify this exercise.
GW propagation in a \qtyproduct{1200x14.2}{\milli\meter} HCSS with surface mounted PZTs of dimensions \qtyproduct{20x0.7}{mm} was analyzed using FE simulations (COMSOL Multiphysics $5.5$). 
The simulation model was set up in a manner similar to the method described in our earlier work \cite{raja2023effect}. A brief overview of the method is presented here. A stationary study was first carried out where the temperature of the HCSS panel was gradually increased from initial room temperature to the desired set-point using \textit{Thermal Expansion} subnode in material properties, and temperature-dependent material properties for the HCSS were derived. Next, time-dependent study was carried out to simulate GW propagation in HCSS at elevated temperature, using \textit{Structural Mechanics} module coupled with \textit{Piezoelectric} interaction physics analyzed using time-implicit study. 
Mapped quadrilateral mesh elements were used in the FE model, with maximum and minimum element sizes of \SI{1}{mm} and \SI{0.01}{mm}, respectively. At \SI{75}{kHz} actuation frequency, the group velocity of the A$0$ mode (the mode employed in this work) is \SI{1.061}{km/s}, and the wavelength is calculated to be \SI{14.44}{mm}. Accurate simulation of Lamb wave propagation in implicit analysis requires at least ten mesh elements per wavelength, and this criterion is thus satisfied. Low reflecting boundary condition was specified at the outer edges of HCSS to reduce reflections from edges, and Rayleigh damping was added for the composite face sheet and the core to account for damping of the GWs during propagation.

The FE model contained \SI{81040}{} elements with \SI{498604}{} degrees of freedom. The simulation time step was chosen to be \SI{25}{\nano\second} for $5-$cycle Hanning pulse actuation centered at \SI{75}{kHz}, for total simulation time of \SI{400}{\micro\second}. The computation time for each simulation on $64-$bit Intel\textregistered~Core\texttrademark-i$7-10$ $750$ H \SI{2.60}{GHz} CPU workstation with \SI{16}{GB} RAM was \SI{4.15}{\hour}. Simulations for TRF and LFA damages were conducted with the simulated damage with conditions similar to the experiment.
Data were generated for various sizes of damage: \SI{5}{\milli\meter}, \SI{10}{\milli\meter}, \SI{15}{\milli\meter}, and \SI{20}{\milli\meter} and at various temperatures in the range of \qtyrange{30}{90}{\degreeCelsius} in intervals of \SI{10}{\degreeCelsius}.
The data were added with white and pink noise, in a manner similar to the noise-augmentation performed for experimental data to obtain SNR of \SI{20}{dB}. 

\subsection{Feature extraction}
In any damage detection approach, choosing the right features from GW signals is a crucial step since feature selection greatly affects accuracy of the damage assessment algorithm. Automated feature extraction using CNNs require training large networks that may not be feasible for edge implementation. While frequency-domain methods such as frequency decomposition  and time-frequency methods such as spectrogram (obtained using short time Fourier transform) and scalogram (obtained using wavelet transform) are best suited to capture the non-stationary nature of GW signals \cite{raghavan2007guided,torbol2014real}, they require substantial computation resources and are not feasible for embedded system implementation for edge-inference \cite{asutkar2023tinyml}.
Features computed from time-domain signals acquired using the embedded system are therefore best suited for implementing ML models for damage inference at the edge. The utility of easy-to-compute statistical metrics obtained from time-domain signals for machine fault diagnosis was shown by Bandyopadhyay et al. \cite{bandyopadhyay2018performance}. 
For the damage assessment model presented in this work, we utilized $16$ features computed from the time-domain signals, summarized in Table \ref{tab:time_features}. Along with widely used statistical features such as mean absolute deviation (MAD), variance, standard deviation, root mean square energy (RMS), root mean square deviation (RMSD), kurtosis, skew, crest factor, impulse factor, shape factor and difference in peak to peak amplitude of signals, more advanced features based on signal energy \cite{Torkamani_2014} were included to capture the impact of damage on amplitude and phase of the GW signals. MAD, variance, and standard deviation measure the deviation of the guided wave signal relative to the mean. However, these features only use one signal at a time, which does not allow comparison between two signals obtained for different states of the structure (e.g. healthy and damaged). Therefore, the remaining features were formulated by comparing a reference signal, usually derived from data for healthy structure taken at room temperature, to the signal obtained using the system in presence of damage or temperature variations. 

\begin{table}[!t]
\begin{center}
\centering
\small
\caption{Features utilized for damage assessment. Glossary: $f(i)$: data samples in the signal, $i=1,2,\dotsc,n$, $n$: length of time series, $X$: ordered list of values in the series, $\mu$:mean value of samples in time-series $f$,$\sigma$: standard deviation of samples in time-series $f$, $f_b$: baseline signal, $A$: peak to peak amplitude of signal $f$, $A_b$: peak to peak amplitude of baseline signal $f_b$ } 

\label{tab:time_features}
\begin{tabular}{p{5cm}|p{3cm}}
\hline
\centering \textbf{Feature} & \textbf{Expression} \\ 
\hline 
\hline
\centering Mean & $\frac{1}{n}\sum_{i=1}^{i=n}f(i)$ \\ [1ex]
\centering Median & 
$\begin{cases}
  X[\frac{n+1}{2}], & n\text{ is odd} 
\\ 
  \frac {{X[\frac{n}{2}] + X[\frac{n}{2}+1]}}{2}, & n\text{ is even}
\end{cases}$
\\  [1ex]
\centering Mean absolute deviation & $\frac{1}{n}\sum_{i=1}^{i=n}|f(i) - \mu|$  \\ [1ex]
\centering Variance & $\frac{\sum_{i=1}^{i=n}{(f(i) - \mu)^2}}{n}$ \\ [1ex]
\centering Standard deviation & $\sqrt{\frac{\sum_{i=1}^{i=n}{(f(i) - \mu)^2}}{n}}$ \\ [1ex]
\centering Root mean square (RMS) & $\sqrt{(\frac{1}{n})\int^n[f(i)]^2 dt}$ \\ [1ex]
\centering Root mean square deviation (RMSD) & $\sqrt{\frac{\int^n[f(t)-f_b(t)]^2 dt}{\int^n[f_b(t)]^2dt}}$ \\ [1ex]
\centering Kurtosis & $\frac{1}{n}\sum_{i=1}^{i=n}\left(\frac{x_i - \mu}{\sigma}\right)^4$ \\ [1ex]
\centering Skew & $\frac{3(\mu - \mu_{1/2})}{\sigma}$\\ [1ex]
\centering Crest factor & $\text{max}(|x_i|) \div \sqrt{\frac{\sum_{i=1}^{i=n} x_i^2}{n}}$ \\ [1ex]
\centering Impulse factor & $\text{max}(|x_i|) \div {\frac{\sum_{i=1}^{i=n} |x_i|}{n}}$ \\ [1ex]
\centering Shape factor & $\sqrt{\frac{\sum_{i=1}^{i=n} x_i^2}{n}} \div \frac{\sum_{i=1}^{i=n} |x_i|}{n}$ \\ [1ex]
\centering Peak to peak & $A-A_b$ \\ [1ex]
\centering Ratio of signal energy & ${\frac{\int^T{|f(t)|^2}dt}{\int^T{|f_{b}(t)|^2}dt}}$ \\ [1ex]
\centering Damage Index  &${\frac{\int^T{|f(t)-f_{b}(t)|^2}dt}{\int^T{|f_{b}(t)|^2}dt}}$\\[1ex]
\centering Normalized difference of signal energy & ${\frac{\int^n{|f(t)|^2}dt-\int^n{|f_{b}(t)|^2}dt}{\int^n{|f_{b}(t)|^2}dt}}$ \\ [1ex]
\hline
\end{tabular}
\end{center}
\end{table}

\subsection{Unsupervised learning}

\begin{figure}[!tbp]
    \centering
    \includegraphics[width=\linewidth]{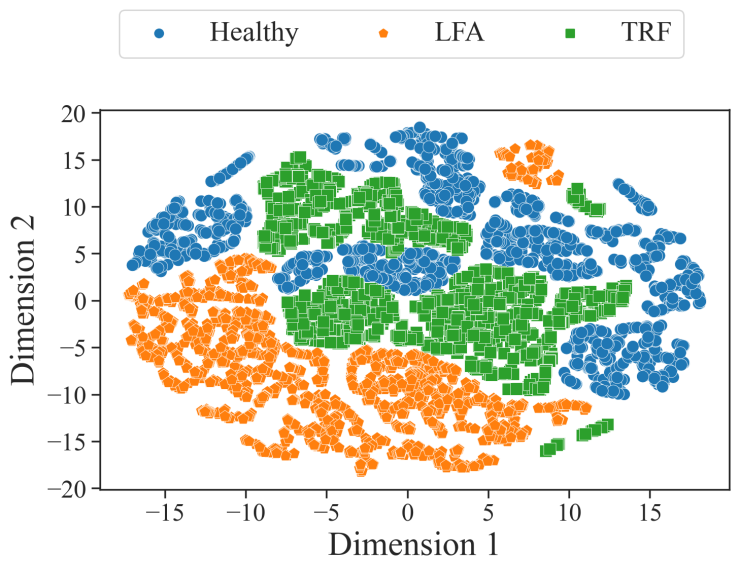}
    \caption{t-SNE plot of experimental data.}
    \label{fig:tsne}
\end{figure}

Variations in the experimental data were studied using T-distributed stochastic neighbor embedding (t-SNE) technique in two dimensions, as shown in Figure \ref{fig:tsne}.
The figure shows clusters belonging to different classes, with each data recording represented using $16$ features listed in Table \ref{tab:time_features}. The features of data obtained from damaged and undamaged paths are not linearly separable, and therefore an ML model is suitable for distinguishing damaged and healthy operating conditions. 
Generating extensive labeled data to capture all possible damage scenarios under all possible varying environmental and operating conditions is impractical, and thus significantly limits the applications of supervised learning algorithms for GW-SHM. Therefore, we have explored the use of unsupervised learning methods to overcome these limitations. One such approach is anomaly detection, which involves finding outliers in data that were not represented in the training data, and mark them as possible damage scenarios. This involves a neural network that is trained for an approximate reproduction or reconstruction of the input signal by passing it through a series of neural layers.
This method is different from supervised learning, because there are no separate discrete (classification) or continuous (regression) labels associated with input data. Training an unsupervised learning model using only data obtained for baseline operation (healthy or undamaged condition) at various temperatures would enable the model to learn variations due to temperature. This would in turn be used to distinguish the same from variations due to damages, which would be identified as outliers, since data from damaged conditions would not be part of the training exercise.
Anomalies in the data due to damage will result in higher error in reconstruction, and therefore be identified as data belonging to damaged condition. The reconstruction error is quantified by the mean square error (MSE) between reconstructed and input data:

\begin{equation}
    MSE = {\frac{1}{n}} \times {\sum_{j=1}^{j=n}{\left(a_{j} - \bar{a_{j}}\right)^2}}
    \label{equation:MSE}
\end{equation}
where $n$ denotes number of elements in the input data, ${a_{j}}$ is the $j^{th}$ element in the input data and ${\bar{a_{j}}}$ is $j^{th}$ element in the reconstructed output data.
The accuracy of the model is calculated by setting the reconstruction error threshold for determining anomaly as $\mu + \sigma$, where $\mu$ and $\sigma$ are the mean and standard deviation respectively of the distribution of reconstruction error obtained from training with healthy data.

\section{Results and discussion}
Instead of setting up a separate model for learning features from vertical and horizontal paths, a single model was trained on combined data from horizontal and vertical paths, to reduce the computational burden on the embedded edge platform. The performance of the model was evaluated for experimental and simulation datasets. To calculate the $16$ features from time-series data, segments corresponding to the initial \SI{200}{\micro\second} portion of the time-series were utilized. Note that the A$0$ mode is expected to be most sensitive to damage. However, we have also included portions corresponding to other GW modes, so that a generic model may be developed, that would not require prior knowledge of material properties of the structure under test, for determination of group velocity and thereby arrival time of a desired GW mode.

\subsection{Analysis of experimental data}
\label{sec:expt computer}
\begin{figure}[!t]
    \centering
    \includegraphics[width=0.9\linewidth]{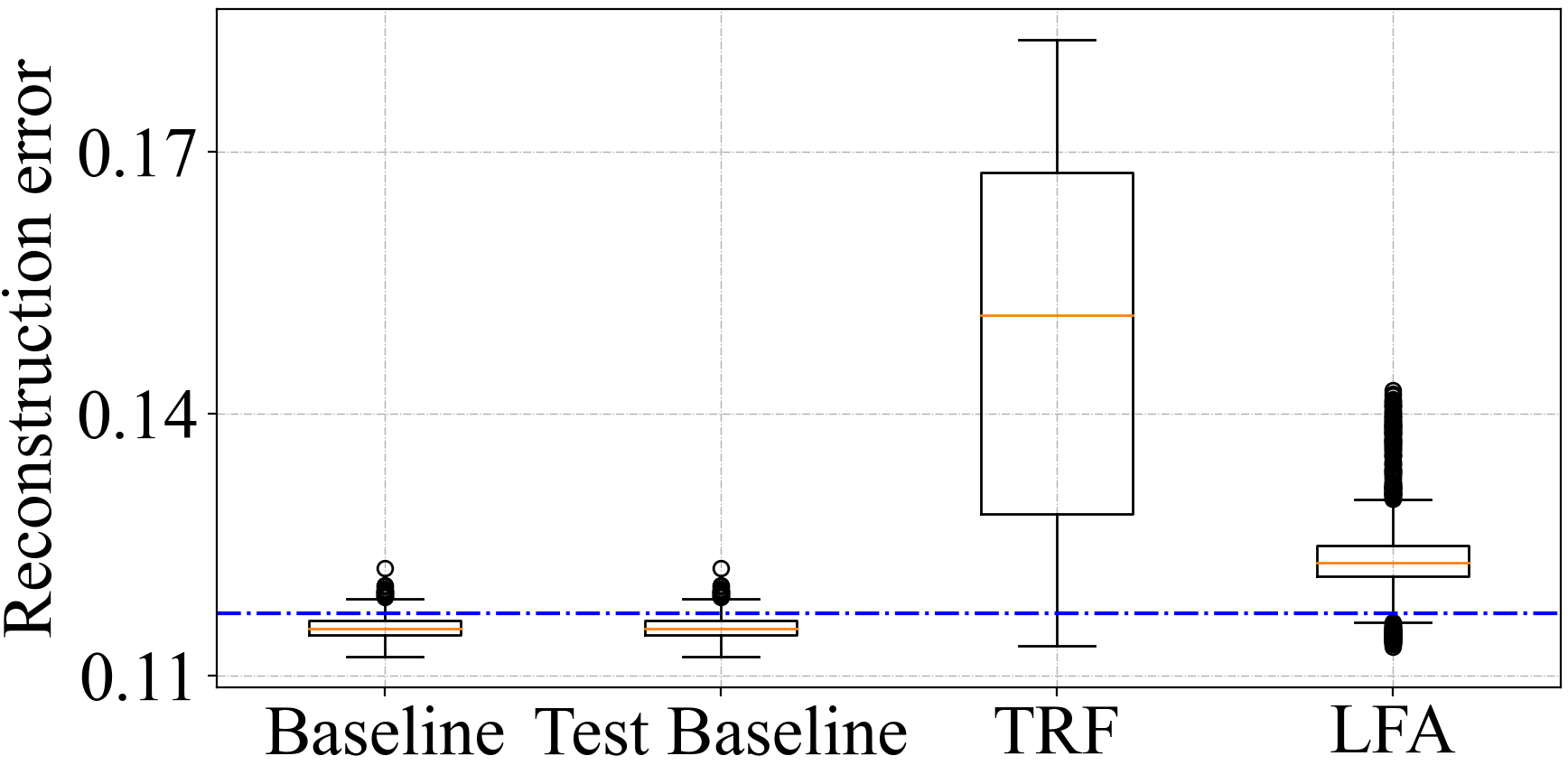}
    \caption{Reconstruction error for experimentally generated data}
    \label{fig:expt_result}
\end{figure}

Experimental data collected on undamaged panel were used for training the model, which was then tested on data collected on panel with TRF and LFA damage. A total of $50 \times 19 \times 6 =  5700$ time-series recordings were available for each of baseline, TRF and LFA conditions: $50$ noise augmented copies generated at $19$ temperatures varying from \qtyrange{0}{90}{\degreeCelsius} for $6$ paths.
For damage assessment, we implemented a neural network with only fully-connected layers, in order to reduce the number of trainable parameters, and therefore resource utilization on the FPGA device. All data were normalized such that all values lie in the range of $[-1,1]$, and the normalized data were used for noise augmentation and feature extraction. The model architecture is summarized in Table \ref{tab:model arch}. The model was developed in Python using TensorFlow library and Keras environment. The network was trained using rectified linear unit (ReLU) activation function and Adam optimizer. Keeping the number of neurons fixed, the batch size, learning rate and number of epochs were optimized using RandomSearch algorithm executed for $10$ iterations. The best set of parameters was then selected based on maximum accuracy obtained with $5-$fold cross-validation score. The range of hyper-parameter values passed to the algorithm and optimal values thus obtained are listed in Table \ref{tab:hyperparameter}. Data samples corresponding to undamaged (i.e., baseline) condition were randomly split into $50$\% for training , $20$\% for validation, and the remaining $30$\% for testing the model. After training, we tested the model performance on another independently collected dataset consisting of data from undamaged, as well as damaged portions. Figure \ref{fig:expt_result} shows distribution of reconstruction error for model predictions for these experimentally generated data for test data of the baseline set used for training, and the independently recorded dataset for testing the model, comprising of baseline, TRF and LFA conditions. The paths that do not contain damage in-line (i.e., \textit{baseline} and \textit{test baseline}) are correctly identified, as the reconstruction error is lesser than the $\mu + \sigma$ threshold. Similarly, damaged paths TRF and LFA are also identified correctly, as the reconstruction error is greater than the $\mu + \sigma$ threshold. Accuracy and F1 score for these results are shown in Table \ref{tab:expt_acc}. 
These results suggest that the chosen set of input features are suitable for capturing the variation in data due to temperature, while simultaneously enabling the model to distinguish changes due to presence of damage.

\begin{table}[!t]
\begin{center}
\centering
\caption{Summary of model architecture}
\label{tab:model arch}
\begin{tabular}{c|c}
\hline
\textbf{Layer (\# filters)}                               & \textbf{\# Parameters} \\ \hline \hline
Dense ($16$)                                      & $272$        \\ \hline
Dense ($32$)                                      & $544$        \\ \hline
Dense ($64$)                                      & $2112$       \\ \hline
Dense ($64$)                                      & $0$          \\ \hline
Dense ($64$)                                      & $4160$       \\ \hline
Dense ($32$)                                      & $2080$       \\ \hline
Dense ($16$)                                      & $528$        \\ \hline
\multicolumn{1}{l|}{Total trainable parameters} & $9696$      \\ \hline
\end{tabular}
\end{center}
\end{table}

\begin{table}[!t]
\begin{center}
\centering
\caption{Hyperparameter tuning using RandomSearch}
\label{tab:hyperparameter}
\begin{tabular}{c|c|c}
\hline
\textbf{Hyperparameter} & \textbf{Range}                   & \textbf{Optimal value} \\ \hline \hline
Learning rate  & {[}$0.001$, $0.01$, $0.1${]}  & $0.01$          \\ \hline
Batch size     & {[}$16$, $28$, $32$, $64${]}    & $32$            \\ \hline
No. of epochs  & {[}$50$, $100$, $150$, $200${]} & $150$           \\ \hline
\end{tabular}
\end{center}
\end{table}

\begin{table}[!tbp]
\begin{center}
\centering
\caption{Accuracy and F1 score for experimental data}
\label{tab:expt_acc}
\begin{tabular}{c|c|c}
\hline
\textbf{Case} & \textbf{Accuracy (\%)} & \textbf{F1 score (\%)} \\ \hline \hline
Test Baseline  & $99$                            & $99$                            \\ \hline
TRF            & $95$                            & $86$                            \\ \hline
LFA            & $95$                            & $83$                            \\ \hline
\end{tabular}
\end{center}
\end{table}

\subsection{Analysis of simulation data}
To study the performance of the damage identification for temperature-affected and noise-augmented data corresponding to various damage sizes, FE simulation data were generated using COMSOL Multiphysics (with noise augmentation performed using MATLAB). A total of $50 \times 7 \times 3 =1050$ datasets were available for training and validation of the model: $50$ noise augmented copies were generated for each of the $7$ temperatures varying from \qtyrange{30}{90}{\degreeCelsius} for $3$ paths.
The same model architecture as used for evaluating experimental data was used for analyzing these simulation generated data, with the only exception being the batch size, which was set as $28$, since we observed the model to over-fit to the data for batch size of $32$.

\begin{figure}[!t]
    \centering
    \includegraphics[width=0.9\linewidth]{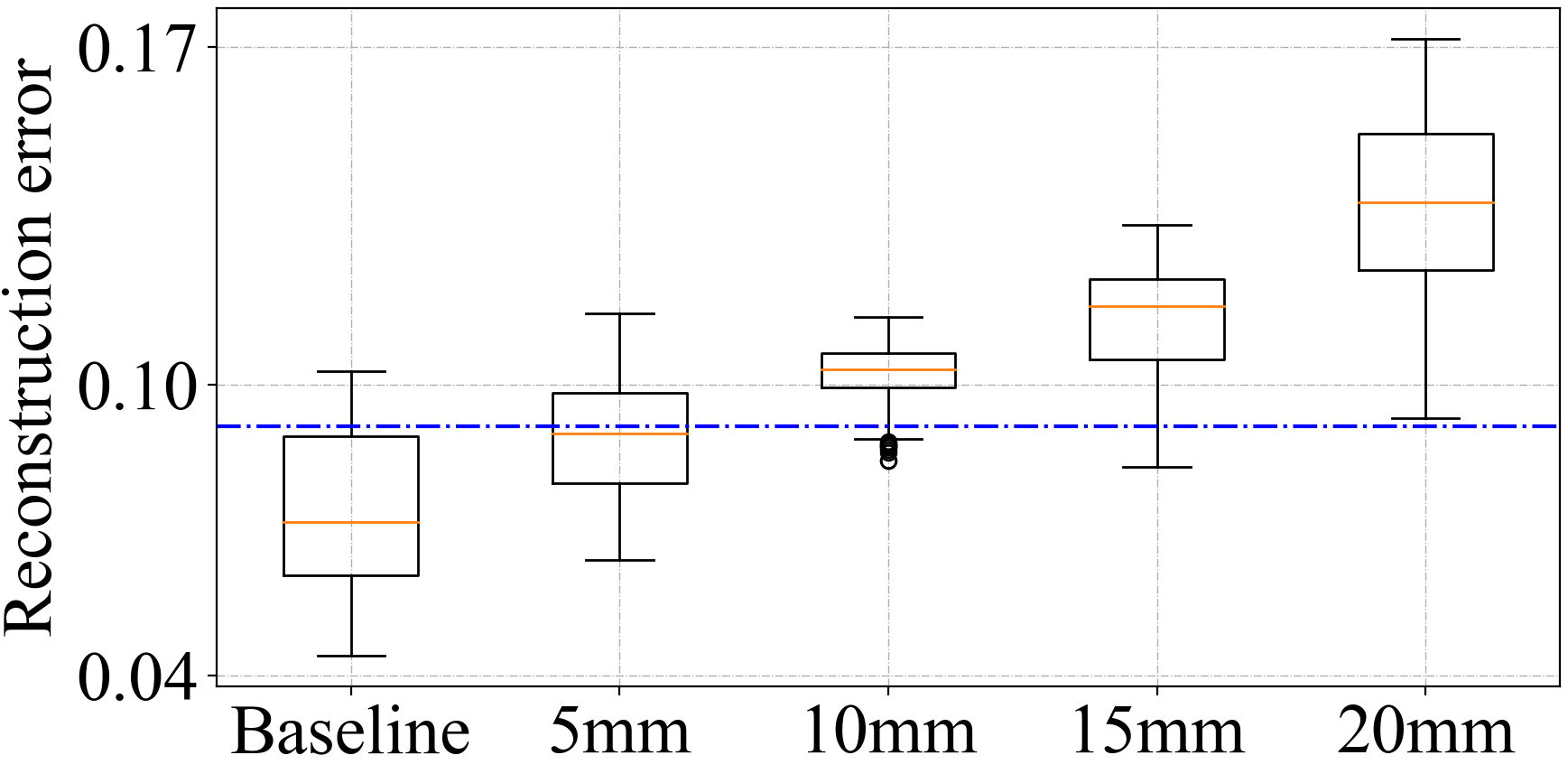}
    \caption{Reconstruction error for TRF data generated by FE simulations}
    \label{fig:TRF sim}
\end{figure}

\begin{figure}[!t]
    \centering
    \includegraphics[width=0.9\linewidth]{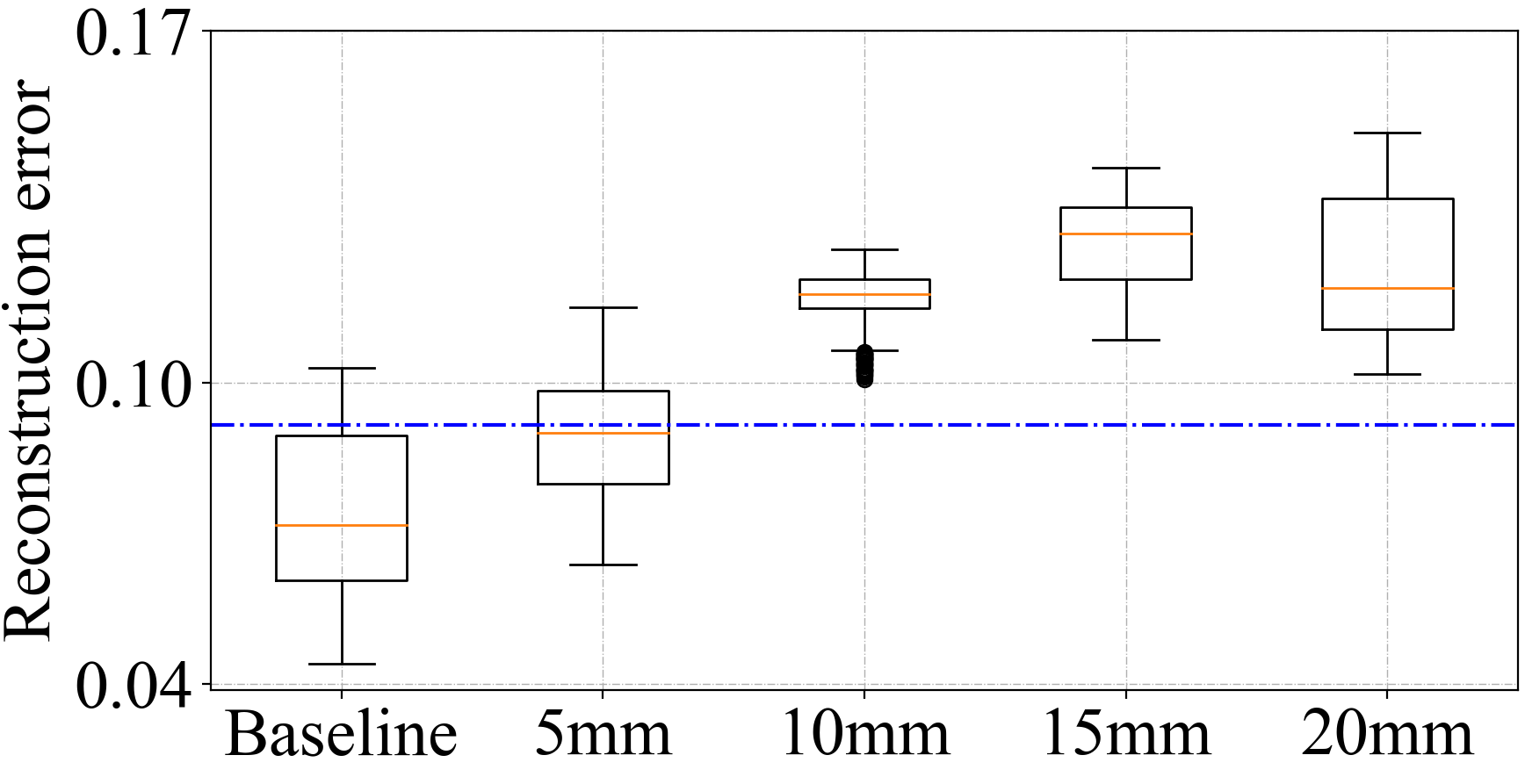}
    \caption{Reconstruction error for LFA data generated by FE simulations}
    \label{fig:LFA sim}
\end{figure}

Figures \ref{fig:TRF sim} and \ref{fig:LFA sim} show distributions of reconstruction error for model predictions for defects of various sizes corresponding to TRF and LFA, respectively. The corresponding accuracy and F1 score metrics are reported in Table \ref{tab:sim_acc}.
Since defects of larger sizes produce larger changes in signal amplitude, the model is able to easily discern TRF and LFA damages of sizes \SI{10}{mm} and larger. The increasing trend of reconstruction error with defect size indicates that the absolute value of reconstruction error could possibly be used as an indicator of defect size (i.e., damage intensity).
Defects of size \SI{5}{mm} are not as accurately classified as damaged. Note that the sensitivity of the model to smaller damages could potentially be improved by exploring transduction at higher frequencies. However, the PZT transducers used in this study were found to have degradation in transduction efficiency at frequencies higher than \SI{150}{kHz}, and therefore only \SI{75}{kHz} actuation was explored in this work. 

\begin{table}[!t]
\begin{center}
\centering
\caption{Accuracy and F1 score for simulation generated data}
\label{tab:sim_acc}
\begin{tabular}{c|cc|cc}
\hline
\textbf{Damage} & \multicolumn{2}{c|}{\textbf{TRF}}                                                  & \multicolumn{2}{c}{\textbf{LFA}}                                                                       \\ \cline{2-5} 
\textbf{size} & \multicolumn{1}{c|}{\textbf{Accuracy}} & \textbf{F1 score} & \multicolumn{1}{c|}{\textbf{Accuracy}} & \multicolumn{1}{c}{\textbf{F1 score}} \\ \hline \hline
\SI{5}{mm}                             & \multicolumn{1}{c|}{$50$}                            & $35$                            & \multicolumn{1}{c|}{$50$}                            & $35$                                                \\ \hline
\SI{10}{mm}                           & \multicolumn{1}{c|}{$90$}                            & $73$                            & \multicolumn{1}{c|}{$96$}                            & $88$                                                \\ \hline
\SI{15}{mm}                            & \multicolumn{1}{c|}{$90$}                            & $73$                            & \multicolumn{1}{c|}{$96$}                            & $88$                                                \\ \hline
\SI{20}{mm}       & \multicolumn{1}{c|}{$96$}                            & $88$                            & \multicolumn{1}{c|}{$96$}                            & $88$                                                \\ \hline
\end{tabular}
\end{center}
\end{table}

\begin{table}[!t]
\begin{center}
\centering
\caption{Resource utilization for model implementation on FPGA device for experimental data}
\label{tab:fpga_resource_utilize}
\begin{tabular}{c|c|c|c}
\hline
\textbf{Resource} & \textbf{Utilized} & \textbf{Available} & \textbf{Utilization (\%)} \\ \hline \hline
LUT      & $5463$        & \SI{20800}{}     & $26.26$          \\ \hline
FF       & $4352$        & \SI{41600}{}     & $10.46$          \\ \hline
BRAM     & $49$          & $50$        & $98$          \\ \hline
DSP      & $5$           & $90$        & $5.56$           \\ \hline
\end{tabular}
\end{center}
\end{table}

\begin{figure}[!t]
    \centering
    \includegraphics[width=0.9\linewidth]{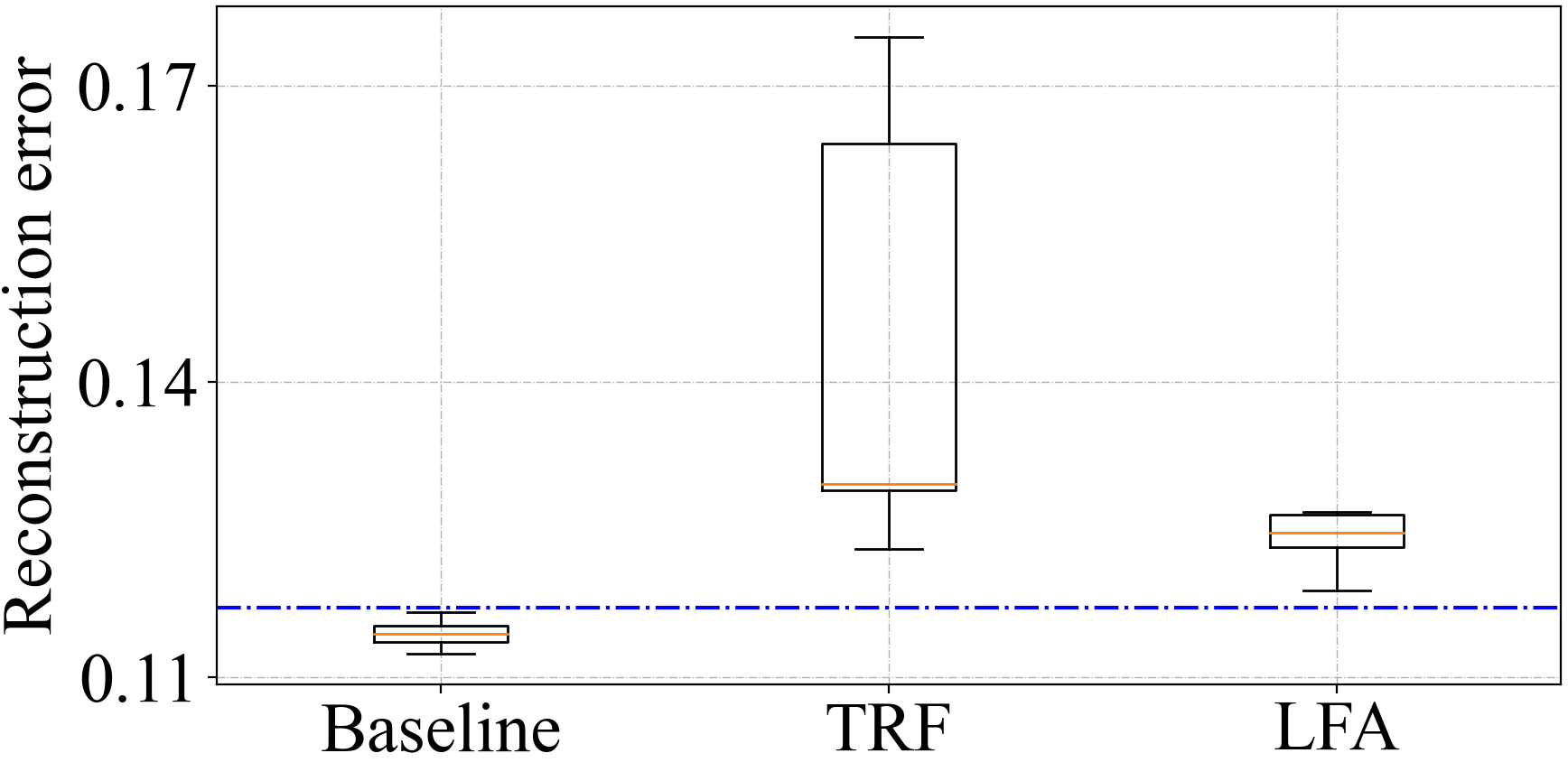}
    \caption{Reconstruction error for experimental data for edge implementation of model}
    \label{fig:fpga_result}
\end{figure}

\begin{table*}[!t]
\begin{center}
\centering
\small
\caption{Comparison of this work with other unsupervised learning algorithms for GW-SHM reported in literature}

\label{tab:comparison}
\begin{tabular}{p{2cm}|p{4.5cm}|p{3.5cm}|p{3cm}|p{4cm}}
\hline
\textbf{Reference }  & \textbf{Application} & \textbf{Input} & \textbf{Architecture} & \textbf{Model size and edge-compatibility}\\ 
\hline 

Lee et al. 
\cite{lee2022automated} & Fatigue damage detection and classification in composite structures & Time-domain signals 
& 1D deep AE & Model size not reported; model not edge-compatible \\[1ex]

Sawant et al. \cite{sawant2023unsupervised} & Damage identification and localization in OGW dataset & Time-domain signals & 1D CAE with transfer learning & $1.1$M parameters; model not edge-compatible \\[1ex]

Eybpoosh et al. 
\cite{eybpoosh2017energy} & Online damage detection of pipelines under varying environmental and operational conditions & Energy-based sparse representation of GW signals & K-means, SVM and Kullback–Leibler divergence (KL) & Model sizes not reported; models not edge-compatible \\ [1ex]

Zhang et al. 
\cite{zhang2020machine} & Automated damage severity and orientation detection & Frequency and time domain features in aluminum beam  & SVM &  Model size not mentioned; accuracy drop on prediction of noise augmented data; model not deployed on FPGA \\[1ex]


Rai et al. 
\cite{rai2022transfer} & GW-based semi-supervised damage diagnosis in composite laminates using ResNet autoencoder with transfer learning (ResNet CAE is not edge-compatible) & Time-domain signals & 1D CAE & $\approx 41$k trainable parameters in 1D CAE with residual layer and $\approx 260$k trainable parameters with transfer learning \\ [1ex]



Rautela et al. 
\cite{rautela2022delamination} & Anomaly (delamination) detection in composite panels & Time-frequency representation as 2D images (continuous wavelet transform)  & 2D CAE & $\approx 1.2$M parameters in encoder and $\approx 1$M trainable parameters in decoder \\[1ex]




Yang et al. \cite{YANG2023110473} & Assessment of damage under regular and irregular variations of EOCs in aluminum plate & Time-domain signals & AE & Method incompatible with edge implementation \\[1ex] 

Sikdar et al. \cite{SIKDAR2023116876} & Assessment of breathing-like debonds in lightweight stiffened composite panels &  Time-frequency representation (continuous wavelet transform)  & Deep CNN & Model not edge-compatible \\[1ex]

\textbf{This work} & TinyML enabled temperature-compensated damage detection in HCSS panel implemented on edge device & Features extracted from time-domain and frequency domain data & ANN & $9696$ parameters; model deployed on Cmod A7-15T FPGA module for data acquisition and inference\\[1ex]




\hline
\end{tabular}
\end{center}
\end{table*}

\subsection{Edge implementation of trained model}
The trained model was deployed using TensorFlow Lite framework on the MicroBlaze® RISC processor core in the Xilinx Artix®-7 device (Digilent Cmod A7-35T module). The challenge in implementing machine learning model in such a device is the absence of a processor core which is required to perform complex control tasks and run TFLite. The MicroBlaze\textregistered~CPU is a family of highly configurable, drop-in, modifiable preset $32-$bit Harvard RISC microprocessor architecture optimised for implementation in Xilinx FPGAs. MicroBlaze\textregistered~is fully supported by Eclipse based integrated development environments (IDEs) Xilinx SDK/Vitis, which allow direct creation of an application project for the Cmod A7-15T module and auto-generation of all board support package (BSP) files. The corresponding TFLite kernels were also modified to invoke the custom IPs for deploying the trained model on MicroBlaze\textregistered. The hardware for FPGA is configured and bitstrean is generated using Vivado $2019.1$. Once the bitstream is generated, BSP is created to provide all low level drivers for the hardware using SDK $2019.1$. The TensorFlow Lite model header file generated using Python application programming interface (API) \textit{TensorFlow Lite converter} is imported in the SDK project, and stored as C byte array in read-only program memory on the FPGA device. For more details on the implementation, please see source code available at \cite{tflitemicro}, which describes the use of this framework for handwritten digit classification in MNIST database and \cite{lall2023deep}, which describes the use of this framework for implementing a reinforcement learning model.
The model architecture described in Table \ref{tab:model arch} was deployed on the device and the performance was evaluated on the same test cases used for evaluating model performance on experimental data reported in section \ref{sec:expt computer}. We observed similar accuracy as that reported in Table \ref{tab:expt_acc}, with average inference time approximately \SI{900}{\micro\second} for \SI{100}{MHz} clock frequency. The resource utilization in the FPGA is summarized in Table \ref{tab:fpga_resource_utilize}, and a plot of reconstruction error for edge implementation is shown in Figure \ref{fig:fpga_result}.

\subsection{Discussion}
Limited memory and computational resources available on embedded edge devices make it challenging to design and deploy machine learning tools for GW-SHM, as arbitrarily large models with significant amount of signal pre-processing typically presented in literature are rendered impractical. This is especially true for permanently deployed SHM systems, as compared to solutions for traditional periodic inspection and nondestructive testing and evaluation (NDT\&E). The application of TinyML for GW-SHM presented in this work illustrates the potential for further exploration towards implementation of autonomous damage inspection strategy on resource constrained hardware. No knowledge of material properties of the structure or impact of damage on guided wave propagation is necessary to apply this method, and therefore it is also potentially scalable to various types of damages in other structures. However, this would require careful examination of damage-sensitive features that must be calculated from the time-series data and provided as input to the model.
Generating the $16$ hand-crafted features from experimentally acquired data reported in this work, and computing reconstruction error takes only a few milliseconds. While the performance can be optimized further using a more capable albeit expensive FPGA device, the Cmod A7-15T module offers an excellent trade-off between performance and price ($<$USD $\$ 100$).
A comparison of our method to other unsupervised learning based damage assessment methods in GW-SHM reported in literature is presented in Table \ref{tab:comparison}. Compared to many of these implementations, our model is extremely light-weight ($<$\SI{10000}{} trainable parameters) and can therefore be effectively stored and executed on the edge device used for configuration and transduction of GW signals. The model inference time on the edge device is $<$\SI{1}{ms}, thereby enabling efficient duty-cycling of such an SHM solution, helping conserve power and battery life in the field. Choosing suitable features allows such a light-weight model to distinguish changes in data due to damage from those due to temperature variation. Thus, our work presents an important breakthrough of relevance to the GW-SHM community for realizing practically relevant smart-sensor nodes that may be permanently deployed on infrastructure.

\section{Conclusion}
In summary, we present a lightweight unsupervised learning algorithm framework that is deployed on an FPGA device for end-to-end implementation of a GW-SHM embedded system for data acquisition, storage, feature extraction and damage assessment. The variance in peak signal amplitude, variation in signal energy, and variation in signal from baseline are all captured by the $16$ hand-crafted features used to represent the time-domain data. These features are adaptable to a wide range of GW-SHM applications because they capture the typical impact of damages on guided wave propagation in structures. The method was tested using data generated from FE simulations on HCSS panels equipped with PZT-5H transducers and an experimental setup for inducing temperature change in the range of \qtyrange{0}{90}{\degreeCelsius}. The unsupervised machine learning model used for damage assessment consists of only $9696$ parameters, and shows high accuracy when trained and tested with noise augmented data with \SI{20}{dB} SNR. The trained model was also successfully deployed on the  MicroBlaze\textregistered~RISC processor core in the Xilinx Artix\textregistered-7 FPGA device. Despite the limited number of features and small model size employed for edge implementation, the models evaluated in this work showed reasonably high accuracy for disbond and delamination defects. We aim to investigate feasibility of the method proposed in this work for different types of structures, damages and severity. In future work, we shall also explore strategies to improve the sensitivity of this framework to defects of smaller sizes and incorporate online training on the edge device to realize completely autonomous self-adaptable GW-SHM systems.

\section*{Acknowledgments}
This work was supported through grants from Science and Engineering Research Board (SERB), Government of India [grant no. CRG/2021/001959] and Indian Space Research Organization (ISRO) [grant no. RD/0118-ISROC00-006]. The authors acknowledge support from staff and access to facilities at the Wadhwani Electronics Lab (WEL), Department of Electrical Engineering, IIT Bombay for carrying out experiments reported in this work. The authors thank Mr. Aryan Lall for initial assistance with deploying TFLite model on Artix\textregistered-7 FPGA.

\section*{Declaration of competing interests}
The authors declare that they have no known competing financial interests or personal relationships that could have appeared to influence the work reported in this paper.

\section*{Data availability}
Data available upon reasonable request.

\bibliography{main}

\end{document}